\documentclass[aps,nofootinbib,preprintnumbers,citeautoscript]{revtex4}
\usepackage{amsmath,amssymb}
\usepackage{enumerate}
\usepackage{graphicx}
\usepackage{amsthm}
\usepackage{mathtools}
\usepackage{textcomp}
\usepackage{braket}
\usepackage{url}
\usepackage{caption}
\usepackage{lineno}

\begin{document}
	\title{Implementing Entangled States on a Quantum Computer}

	\author{Amandeep Singh Bhatia$^1$ and Mandeep Kaur Saggi$^2$\\
		\textit{$^1$Department of Computer Science, Chitkara University, India} \\
		\textit{$^2$Department of Computer Science, Thapar Institute of Engineering \& Technology, India} \\
		E-mail: $^1$amandeepbhatia.singh@gmail.com\\
	E-mail: $^2$mandeepsaggi90@gmail.com}

	\begin{abstract}
		The study of tensor network theory is an important field and promises a wide range of  experimental and quantum information theoretical applications. Matrix product state is the most well-known example of tensor network states, which provides an effective and efficient representation of one-dimensional quantum systems. Indeed, it lies at the
		heart of density matrix renormalization group (DMRG), a most common method for simulation of one-dimensional strongly correlated quantum systems. It has got attention from several areas varying from solid-state systems to quantum computing and quantum simulators.  We have considered maximally entangled matrix product states (GHZ and W).  Here, we designed 
		the quantum circuits for implementing the matrix product states. In this paper, we simulated the matrix product states in customized IBM’s (2-qubit, 3-qubit and 4-qubit) quantum systems and determined the probability distribution among the quantum states.
	\end{abstract}
	\maketitle

	

	\theoremstyle{plain}
	
	\newtheorem{thm}{Theorem}
	
	\theoremstyle{definition}
	\newtheorem{defn}{Definition}
	\newtheorem{exmp}{Example}

	\section{Introduction and motivation}
	Since, Feynman \cite{1} proposed the idea of quantum computing and stated that quantum computers
	can simulate quantum mechanical systems exponentially faster, outstanding advancement has been made in  simulation of quantum systems. In the last decade, the quantum simulation of closed and open systems has got overwhelming response among research communities. It promises powerful applications in the field of high energy physics, quantum chemistry and condensed matter, which are intractable on classical computers. The field of quantum computing is concern with the behavior and nature of energy at the quantum level to improve the efficiency of computations. The main aim of running quantum algorithms on quantum computers to solve the various computational tasks more efficiently and in less time as compared to existing classical ones. The quantum principle of superposition and entanglement is the backbone of quantum algorithms, which allows us to perform operations in large Hilbert spaces exponentially.

	Tensor network states are a new language for quantum many-body systems based on pure quantum mechanical phenomena 'entanglement' \cite{2}. Tensor network states are classified on the basis of dimensions along which the tensors are crossed. Thus, it manages the exponentially growing Hilbert space by restricting the entanglement between two parts of the quantum system \cite{4}. Although, the dimension of Hilbert space increases exponentially with the increase in size of the system, there exists some quantum many-body systems whose implementations are realizable by classical computers. The density matrix renormalization group (DMRG) has been applied in lattice field theory to study particle physics with Monte Carlo simulations \cite{33, 34}. A matrix product state (MPS) (one-dimensional system) is an example of tractable quantum systems numerically. In MPS, low entangled states are efficiently represented, which is not with large dimensions tensor network states. It has some basic properties such as dense nature, finite correlation, translational invariance and one-dimensional area law \cite{3}.

	In the last decade, the simulations of time dependent quantum states have been widely used in different physical systems with MPS \cite{43, 44, 45, 46}. There are several softwares consisting high performance libraries and parameters build upon tensor network theory such as C++ library (ZKCM\_QC) \cite{37} for multi-precision, QCMaquis \cite{38} for optimization purpose, EvoMPS \cite{36} for simulating time-dependent (real or imaginary) one-dimensional many-particles, DMRG++ \cite{39} an open source implementation of DMRG algorithm, iTensor \cite{40} a C++ library for executing tensor network algorithms, Uni10 \cite{41} an open-source and free C++ library for the construction of tensor network algorithms, QuTiP \cite{20} open source python framework, OSMPS-open-source MPS \cite{42}, Symbolic C++ quantum simulation \cite{47} and many more.

In the past few years, the research has been increased in simulating experiments on IBM (International Business Machines Corporation) Q Experience platform. IBM has given access to real quantum computers and simulators, which allows researchers to develop, test and
implement their experiments \cite{5}. Through IBM, we can examined the simulation results on classical computer and analysis on available quantum hardware to get a feel of the quantum system. It is becoming far more widespread and efficiently used to simulate several computational problems quickly \cite{6}. Currently, the platform has been extensively used to implement several experiments such as  quantum tunneling simulation \cite{7}, Ising model simulation hard problems \cite{9}, quantum algorithms \cite{10}, quantum error correction \cite{11}, quantum machine learning \cite{14}, benchmarking the quantum gates \cite{28} to name a few.

	Recently, matrix product state (MPS) provides a stepping stone to various advancements in field of condensed matter and quantum computational theory: unsupervised learning using MPS \cite{31}, quantum dynamics \cite{17, 19}, open source MPS simulation \cite{20}, simulating quantum computation \cite{22}, simulation of open quantum systems \cite{18}, quantum finite state machines of MPS \cite{16}, supervised learning \cite{23} and neural network representation \cite{24} and MPS based efficient, productive applications for high performance computers \cite{29} and can be employed in various emerging technologies such as optical computing, quantum cryptography, image recognition and dynamic quantum clustering \cite{30}. MPS provides an efficient approximation of realistic local Hamiltonians and can be generated by tensors sequentially. We focused on the maximum entangled MPS (GHZ and W states). Zizzi \cite{32} has shown that it is possible to preserve the probabilities by executing a reversible quantum measurement on closed systems. In previous paper, we have efficiently simulated MPS with a broader quantum computational theory and investigated their relationship with quantum finite-state machine (QFSM) using unitary criteria \cite{16}. In this paper, we have implemented simulation on customized real IBM quantum computer and simulator. Further, the probability distributions among qubits is investigated. The paper is organized as follows: Section 2 is devoted to the family of matrix product state. In Section 3, simulation results are presented. Finally, Section 4 is the conclusion.

	\section{Matrix product state}

	Matrix product state is complete. Basically, it concede the extent of entanglement in bond dimensions. In fact, any pure quantum state can be described by substituting the coefficients e.g. rank-\textit{N} tensor by \textit{N}-rank 3 tensors and 2-rank by 2 tensors. In MPS, a pure quantum state $\ket{\phi}$ is represented as:
	\begin{equation}
	\ket{\phi}=\sum_{\sigma_1, \sigma_2,...\sigma_L}^{d} Tr[M_1^{\sigma_1} M_2^{\sigma_2}...M_L^{\sigma_L}] \ket{\sigma_1, \sigma_2,...\sigma_L}
	\end{equation}
	where $M_i^{\sigma_i}$ are complex square matrices, \textit{d} is dimension, $\sigma_i$ represents the indices i.e. \{0, 1\} for qubits and \textit{Tr}() denotes trace of matrices \cite{17}.  Figure 1 shows the MPS as one-dimensional array of tensors and an instance of finite system of 5 sites \cite{3}.  The GHZ state and W state can be represented using MPS: 
	
	\begin{figure}[h]
		\centering
		\includegraphics[scale=0.5]{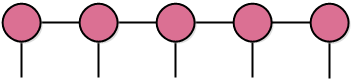}
		\caption{Representation of MPS with 5 sites} \label{1}
	\end{figure}

	\begin{itemize}
		\item \textit{GHZ state}: A Greenberger-Horne-Zeilinger (GHZ) \cite{26} state of \textit{N}-spins 1/2 is represented  as a 
		\begin{equation}
		\ket{GHZ}=\dfrac{1}{\sqrt{2}}(\ket{0}^{\otimes N}+\ket{1}^{\otimes N})
		\end{equation}
		
		 It is defined as maximum entangled state, consisting some non-trivial entanglement features \cite{27}. Pan et al. \cite{48} presented schemes for preparing GHZ state remotely. The 4-qubit GHZ state is represented as \(\frac{1}{\sqrt{2}}\left( {|0000\rangle +|1111\rangle } \right)\). Its matrix product state and unitary matrix (\textit{U}$_{GHZ}$) are represented as:
		 \begin{equation}
		 M^{0}=\begin{bmatrix}
		 1 & 0 \\0 & 0
		 \end{bmatrix}, \quad M^{1}=\begin{bmatrix}
		 0 & 0 \\0 & 1
		 \end{bmatrix}, \quad U_{GHZ}=\frac{1}{\sqrt{2}}\begin{bmatrix}
		 1 & -1 \\1 & 1
		 \end{bmatrix}
		 \end{equation}

		\item \textit{W state}: The \textit{n}-qubit W state is represented as \cite{}:
		\begin{equation}
		\ket{W}=\dfrac{1}{\sqrt{n}}(\ket{100...0}+\ket{010...0}+...+\ket{000...1})
		\end{equation} 
		The W state refers to the superposition of pure entangled states with same coefficients \cite{27}. It is different from above GHZ state. It represents a multipartite entanglement, where one of the qubits is in up state $\ket{1}$, while others are in down state $\ket{0}$. The 4-qubit W state is represented as  \(\frac{1}{\sqrt{4}}\left( {|1000\rangle \;+|0100\rangle \;+\;|0010\rangle  \;+\;|0001\rangle } \right) \) \cite{17}. The matrix representation is given as:
	
		\begin{equation}
		A(1)^{0}= \begin{bmatrix}
		0 & 0 & 0 & 0\\ 0 & 1 & 0 & 0 \\ 0 & 0 & 1 & 0 \\ 0 & 0 & 0 & 1
		\end{bmatrix}, A(2)^{0}=\begin{bmatrix}
		1 & 0 & 0 & 0\\ 0 & 0 & 0 & 0 \\ 0 & 0 & 1 & 0 \\ 0 & 0 & 0 & 1
		\end{bmatrix},  A(3)^{0}=\begin{bmatrix}
		1 & 0 & 0 & 0\\ 0 & 1 & 0 & 0 \\ 0 & 0 & 0 & 0 \\ 0 & 0 & 0 & 1
		\end{bmatrix}, A(4)^{0}=\begin{bmatrix}
		1 & 0 & 0 & 0\\ 0 & 1 & 0 & 0 \\ 0 & 0 & 1 & 0 \\ 0 & 0 & 0 & 0
		\end{bmatrix}
		\end{equation}
		\begin{equation}
		 A(1)^{1}= \begin{bmatrix}
		1 & 0 & 0 & 0\\ 0 & 0 & 0 & 0 \\ 0 & 0 & 0 & 0 \\ 0 & 0 & 0 & 0
		\end{bmatrix},  A(2)^{1}=\begin{bmatrix}
		0 & 0 & 0 & 0\\ 0 & 1 & 0 & 0 \\ 0 & 0 & 0 & 0 \\ 0 & 0 & 0 & 0
		\end{bmatrix}, A(3)^{1}=\begin{bmatrix}
		0 & 0 & 0 & 0\\ 0 & 0 & 0 & 0 \\ 0 & 0 & 1 & 0 \\ 0 & 0 & 0 & 0
		\end{bmatrix},  A(4)^{1}=\begin{bmatrix}
		0 & 0 & 0 & 0\\ 0 & 0 & 0 & 0 \\ 0 & 0 & 0 & 0 \\ 0 & 0 & 0 & 1
		\end{bmatrix}
		\end{equation}
	\end{itemize}

	\section{Simulation results}
		
		Before, we proceed to the simulation of matrix product state, it is useful to define the notion of some operators used over single qubit in quantum computation. The controlled NOT operator (CNOT) has two inputs as well as outputs \cite{17}. The flip operation is performed over the target qubit when the first qubit is 1. The quantum circuit for CNOT operator is shown in Fig 3 and its matrix representation is given in Eq. (10). The other highly used quantum operators are  flip operator (X): flips the qubit, identity operator (I): it does nothing i.e. outputs the qubit as it is, Hadamard operator (H): it generates superposition of states with equal probability in computational basis. Their matrices are given in Eq. (7-9) respectively. In quantum circuits, such operators can be represented using quantum gates. The universal set of quantum gates are shown in Fig 2.
			\begin{figure*}[h]
			\centering
			\includegraphics[scale=0.75]{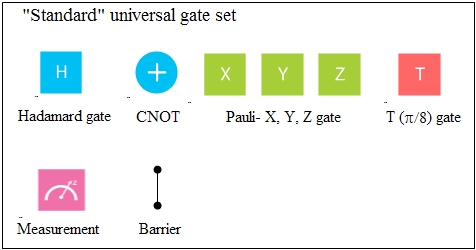}
			\caption{Representation of quantum gates} \label{1}
		\end{figure*}
		\begin{itemize}
	\item \textit{Flip operator} (X): 
	\begin{equation}
	X=\begin{bmatrix}
	0 & 1 \\1 & 0
	\end{bmatrix} \begin{array}{cc}
	X\ket{0}=\ket{1}\\
	X\ket{1}=\ket{0}
	\end{array}
	\end{equation}
	
		\item \textit{Identity operator} (I): 
		\begin{equation}
		I=\begin{bmatrix}
		1 & 0 \\0 & 1
		\end{bmatrix} 
		\begin{array}{cc}
		I\ket{0}=\ket{0}\\
		I\ket{1}=\ket{1}
		\end{array}
		\end{equation}
		\item \textit{Hadamard operator} (H): 
		\begin{equation}
		H=\dfrac{1}{\sqrt{2}}\begin{bmatrix}
		1 & 1 \\1 & -1
		\end{bmatrix} \begin{array}{cc}
		H\ket{0}=\dfrac{1}{\sqrt{2}}(\ket{0}+\ket{1})\\
		H\ket{1}=\dfrac{1}{\sqrt{2}}(\ket{0}-\ket{1})
		\end{array}
		\end{equation}
		\item \textit{CNOT operator}:
		\begin{equation}
		CNOT=\begin{bmatrix}
		1 & 0 & 0 & 0\\
		0 & 1 & 0 & 0\\
		0 & 0 & 0 & 1\\
		0 & 0 & 1 & 0
		\end{bmatrix}
		\end{equation}
		\begin{figure*}[h]
			\centering
			\includegraphics[scale=0.5]{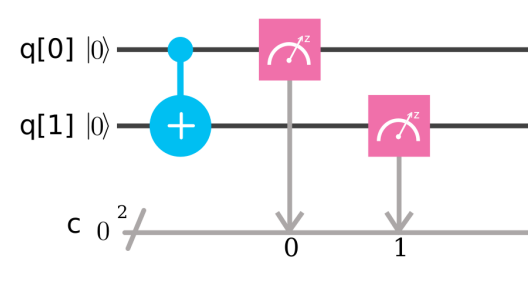}
			\caption{Quantum circuit for CNOT operator} \label{1}
		\end{figure*}
	\end{itemize}

\subsection{GHZ state}
Experimentally, we utilized the IBM Quantum Experience (IBM-QE), a universal customized (2-qubit, 3-qubit and 4-qubit) quantum computer to perform our simulation. The following are the steps to test experiment on IBM-QE:
\begin{itemize}
	\item Construct the quantum circuit for the computational problem through Python programming language and specify it using graphical interface.
	\item Execute the quantum circuit on the IBM simulator and test whether it produces works correctly or not. 
	\item The implementation of the quantum circuit is performed on the hardware processor for given number of shots. During each shot, the specified quantum processor is assigned to implement the quantum circuit. 
	\item Finally, the measurement is performed and the probability of each process is calculated.  
	\end{itemize}

 Here, we started with the maximally entangled 2-qubit GHZ state, which is represented as $\ket{GHZ}_2=\dfrac{1}{\sqrt{2}}(\ket{00}+\ket{11})$. Initially, we have taken two quantum as well as classical registers. Both quantum registers are initialized to zero. The Hadamard gate and CNOT  gate is used to create an entanglement between both qubits, followed by measurement of quantum state q[0] and q[1] to classical registers c[0] and c[1] respectively. Correspondingly, the quantum circuit for 2-qubit GHZ state is shown in Fig. 4. Further, the probability distribution is determined as 53.3906 \% for state $\ket{00}$ and 49.6094 \% for $\ket{11}$ respectively i.e. on summation it satisfies unitary property. The histogram of probability distribution for $\ket{GHZ}_2$ is shown in Fig. 5.

\begin{figure} [h!]
	\centering
	\begin{minipage}{.5\textwidth}
		\centering
		\includegraphics[width=1\linewidth]{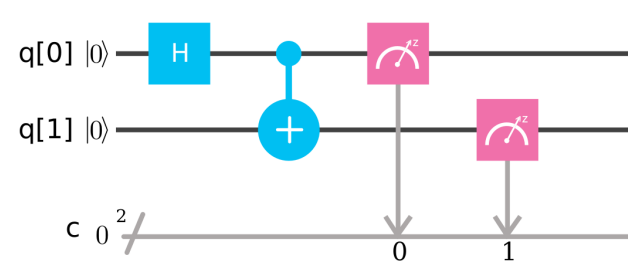}
		\captionof{figure}{Quantum circuit for 2-qubit GHZ state}
		\label{fig:test1}
	\end{minipage}%
	\begin{minipage}{.45\textwidth}
		\centering
		\includegraphics[width=1\linewidth]{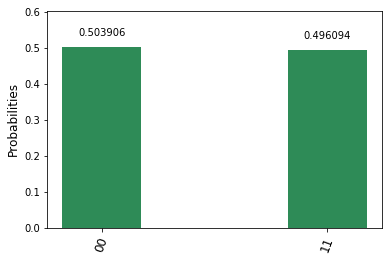}
		\captionof{figure}{Probability distribution for 2-qubit GHZ state}
		\label{fig:test2}
	\end{minipage}
\end{figure}

\begin{figure} [h!]
	\centering
	\begin{minipage}{.5\textwidth}
		\centering
		\includegraphics[width=1\linewidth]{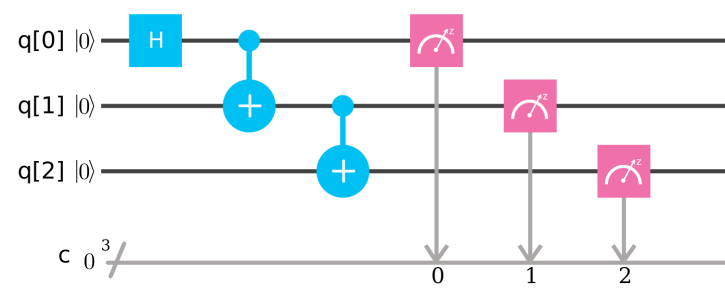}
		\captionof{figure}{Quantum circuit for 3-qubit GHZ state}
		\label{fig:test1}
	\end{minipage}%
	\begin{minipage}{.45\textwidth}
		\centering
		\includegraphics[width=1\linewidth]{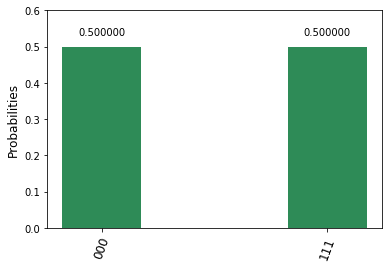}
		\captionof{figure}{Probability distribution for 3-qubit GHZ state}
		\label{fig:test2}
	\end{minipage}
\end{figure}

Next, we have considered 3-qubit GHZ state interpreted  as $\ket{GHZ}_3=\dfrac{1}{\sqrt{2}}(\ket{000}+\ket{111})$. To simulate the $\ket{GHZ}_3$ on quantum computer, we have taken three quantum registers to carry out computation and three classical registers for measurement purpose at the end. The Hadamard and CNOT operation between first and second quantum qubit is similar to $\ket{GHZ}_2$. Then, the second CNOT operation is performed between second and third quantum qubit i.e. flips the q[3] whenever the q[2] is in excited state. The final results can be restirved from classical registers. It creates the 3-qubit entangled GHZ state and  its equivalent quantum circuit is given in Fig 6. Further, the probability distribution is calculated, which comes out to be same i.e. $\dfrac{1}{2}$ for both $\ket{000}$ and $\ket{111}$ respectively.

Further, the complex 4-qubit GHZ quantum state $\ket{GHZ}_4=\dfrac{1}{\sqrt{2}}(\ket{0000}+\ket{1111})$ is taken. A 4-qubit quantum circuit for $\ket{GHZ}_4$ is designed to check whether it is capable of showing quantum behavior on implementing repeatedly. Following the above procedure, the Hadamard gate and CNOT gate produces an entanglement between q[3] and q[4] as shown in Fig 8. Finally, the measurement is readout after one shot and yields the occurrence of state $\ket{0000}$ with 50.2930 \% and $\ket{1111}$ with 49.7070 \% respectively, shown in Fig 9. It can be easily checked that after every shot on quantum simulator, the summation of probabilities of occurrence of both quantum states is comes out to be 1. Thus, it confirms the probability distribution of four spins-1/2 GHZ state into single state i.e. highly entangled quantum state.

\begin{figure*}[h]
	\centering
	\includegraphics[scale=0.55]{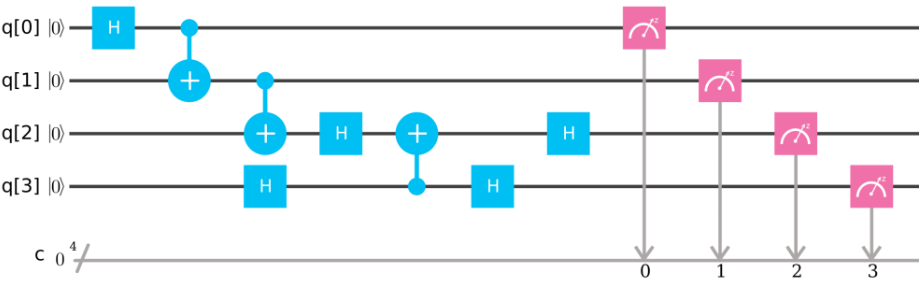}
	\caption{Quantum circuit for 4-qubit GHZ state} \label{1}
\end{figure*}

\begin{figure*}[h]
	\centering
	\includegraphics[scale=0.55]{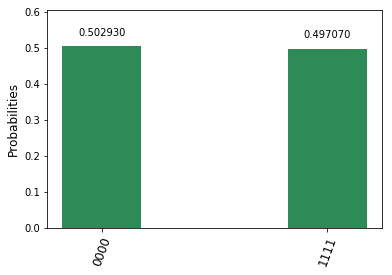}
	\caption{Probability distribution for 4-qubit GHZ state} \label{1}
\end{figure*}

\subsection{W state} 
The GHZ state and W state represent different types of entanglement, where W state is said to be less entangled as compared to GHZ. On measurement, GHZ state is collapse into mixture of other states. But, in case of W state, it leaves the bipartite entanglements on measuring one of its sub-systems.

In this subsection, we have considered 3 and 4-qubit W state and their equivalent quantum circuits are designed and tested on quantum simulator. The 3-qubit W state $\ket{W}_3=\dfrac{1}{\sqrt{3}}(\ket{100}+\ket{010}+\ket{001})$ is considered. Here, we have used flip gate (X) to flip the qubit and Pauli matrix ($R_y$) to rotate the spin in \textit{y}-direction with given theta ($\theta$). After performing the flip operation over the third qubit, barrier is assigned between second $q_1$ and third $q_2$ to avoid the optimization of consecutive gates in a quantum circuit, as shown in Fig 10. Then, CNOT gate is applied successively between $q_0$, $q_1$ and $q_1$, $q_2$ to make one of qubits in excited state and other in ground state same time. 

\begin{figure} [h!]
	\centering
	\begin{minipage}{.5\textwidth}
		\centering
		\includegraphics[width=1\linewidth]{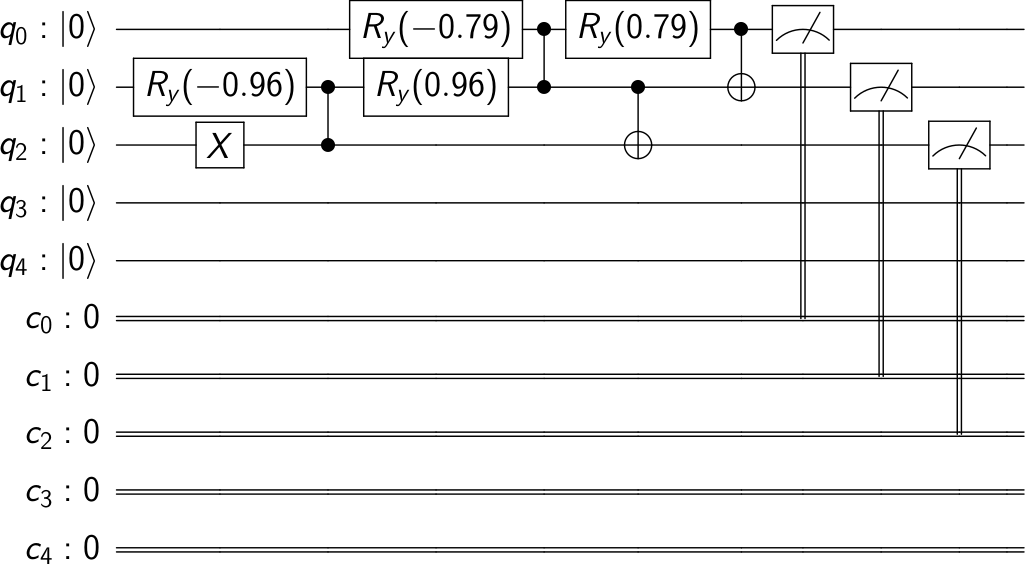}
		\captionof{figure}{Quantum circuit for 3-qubit W state}
		\label{fig:test1}
	\end{minipage}%
	\begin{minipage}{.45\textwidth}
		\centering
		\includegraphics[width=1\linewidth]{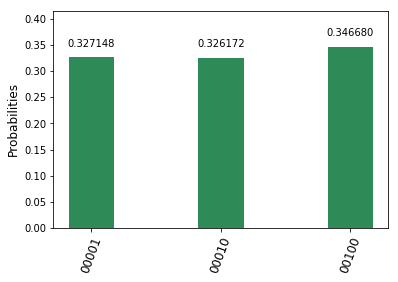}
		\captionof{figure}{Probability distribution for 3-qubit W state}
		\label{fig:test2}
	\end{minipage}
\end{figure}

\begin{figure*}[!h]
	\centering
	\includegraphics[scale=0.32]{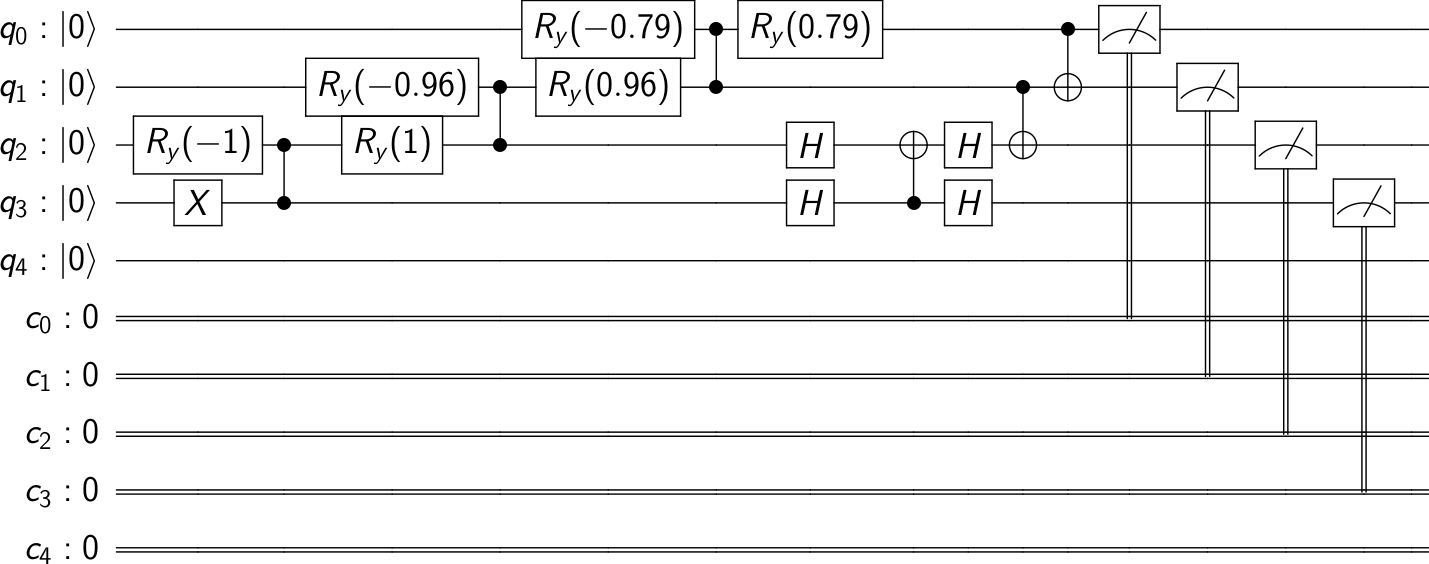}
	\caption{Quantum circuit for 4-qubit W state} \label{1}
\end{figure*}

\begin{figure*}[!h]
	\centering
	\includegraphics[scale=0.62]{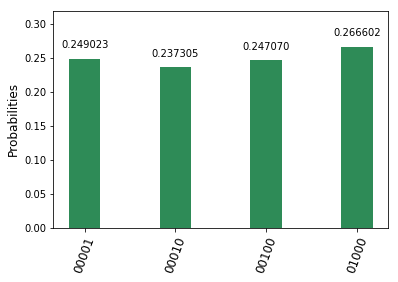}
	\caption{Probability distribution for 4-qubit W state} \label{1}
\end{figure*}

Finally, the equivalent quantum circuit for $\ket{W}_4$ is tested on IBM-QE and results are plotted in Fig. 11. The probability distribution of states is examined as approx 33 \% for each $\ket{100}$ and $\ket{010}$, and 34 \% for $\ket{001}$. It agrees the superimposed 3-qubit W state into singlet state. Furthermore, we constructed the quantum circuit for complicated 4-qubit W state $\ket{W}_4=\dfrac{1}{\sqrt{4}}(\ket{1000}+\ket{0100}+\ket{0010}+\ket{0001})$. Following the above computational procedure of $\ket{W}_3$, we have used combination of Hadamard gates before and after the CNOT operation. Thus, it execute the CNOT operation in reverse direction using the Kronecker product ($H \otimes H$) i.e. equal to $\dfrac{1}{\sqrt{2}}\begin{bmatrix}
H  & H \\H & -H
\end{bmatrix}$ ($4 \times 4$) matrix. The constructed quantum circuit for $\ket{W}_4$ is shown in Fig 12. Finally, after simulation, the measurement is retrieved from classical registers and the histogram of probability distribution among the states is plotted in Fig 13. It shows the occurrence of state $\ket{1000}$ and $\ket{0010}$ with approx 25 \% each, 24 \% of $\ket{0100}$ and state $\ket{0001}$ with 26 \% approximately. Thus, it can be easily checked that \textit{n}-qubit GHZ and W state can be efficiently simulated on a quantum computer after constructing complicated quantum circuits.

\section{Conclusion}
In conclusion, we have demonstrated the two highly entangled matrix product states (GHZ and W). The main focus of this paper is towards the simulation on a quantum computer. We have designed quantum circuits for (2, 3 and 4-qubit) GHZ state and (3 and 4-qubit) W state. Further, the constructed quantum circuits have been simulated on a quantum computer and their probability distribution among the quantum states is investigated. To conclude, we have noticed that any matrix product state can be efficiently simulated on a quantum computer. This allows us to employ these circuits for efficient and productive quantum computational theoretical and experimental applications such as quantum machine learning, MPS based quantum algorithms and can be used in condensed matter physics for quantum state tomography in future. 

\section*{Acknowledgments}
Amandeep Singh Bhatia was supported by Maulana Azad National Fellowship (MANF), funded by Ministry of Minority Affairs, Government of India.

\label{sec:test}
\bibliographystyle{elsarticle-num}
\bibliography{sample}

\begin{thebibliography}{10}
\expandafter\ifx\csname url\endcsname\relax
  \def\url#1{\texttt{#1}}\fi
\expandafter\ifx\csname urlprefix\endcsname\relax\def\urlprefix{URL }\fi
\expandafter\ifx\csname href\endcsname\relax
  \def\href#1#2{#2} \def\path#1{#1}\fi

\bibitem{1}
R.~P. Feynman, Simulating physics with computers, International journal of
  theoretical physics 21~(6) (1982) 467--488.

\bibitem{2}
S.~R. White, Density-matrix algorithms for quantum renormalization groups,
  Physical Review B 48~(14) (1993) 10345.

\bibitem{4}
J.~D. Biamonte, S.~R. Clark, D.~Jaksch, Categorical tensor network states, AIP
  Advances 1~(4) (2011) 042172.

\bibitem{33}
Y.~Shimizu, Tensor renormalization group approach to a lattice boson model,
  Modern Physics Letters A 27~(06) (2012) 1250035.

\bibitem{34}
P.~Zhang, Z.~Xu, H.~Ying, J.~Dai, P.~Crompton, {Quantum Monte Carlo simulations
  of adulteration effect on bond alternating spin= 1/2 chain}, Modern Physics
  Letters A 22~(07n10) (2007) 741--748.

\bibitem{3}
R.~Or{\'u}s, A practical introduction to tensor networks: Matrix product states
  and projected entangled pair states, Annals of Physics 349 (2014) 117--158.

\bibitem{43}
N.~Schuch, M.~M. Wolf, F.~Verstraete, J.~I. Cirac, {Simulation of quantum
  many-body systems with strings of operators and Monte Carlo tensor
  contractions}, Physical review letters 100~(4) (2008) 040501.

\bibitem{44}
G.~Vidal, Efficient simulation of one-dimensional quantum many-body systems,
  Physical review letters 93~(4) (2004) 040502.

\bibitem{45}
M.~Banuls, R.~Or{\'u}s, J.~Latorre, A.~P{\'e}rez, P.~Ruiz-Femenia, Simulation
  of many-qubit quantum computation with matrix product states, Physical Review
  A 73~(2) (2006) 022344.

\bibitem{46}
G.~Vidal, Efficient classical simulation of slightly entangled quantum
  computations, Physical review letters 91~(14) (2003) 147902.

\bibitem{37}
A.~SaiToh, A multiprecision c++ library for matrix-product-state simulation of
  quantum computing: Evaluation of numerical errors, in: Journal of Physics:
  Conference Series, Vol. 454, IOP Publishing, 2013, p. 012064.

\bibitem{38}
L.~Freitag, S.~Keller, S.~Knecht, Y.~Ma, C.~Stein, M.~Reiher, {A quick user
  guide to the QCMaquis software suite for OpenMolcas}.

\bibitem{36}
A.~Milsted, T.~Osborne, {evoMPS}, \url{https://github.com/amilsted/evoMPS}.

\bibitem{39}
G.~Alvarez, The density matrix renormalization group algorithm for strongly
  correlated systems: A generic implementation, Bulletin of the American
  Physical Society 54.

\bibitem{40}
E.~M. Stoudenmire, S.~R. White, {ITensor-Intelligent Tensor Library},
  \url{http://itensor.org/}.

\bibitem{41}
Y.-J. Kao, Y.-D. Hsieh, P.~Chen, {Uni10: An open-source library for tensor
  network algorithms}, in: Journal of Physics: Conference Series, Vol. 640, IOP
  Publishing, 2015, p. 012040.

\bibitem{20}
J.~Johansson, P.~Nation, F.~Nori, {QuTiP: An open-source Python framework for
  the dynamics of open quantum systems}, Computer Physics Communications
  183~(8) (2012) 1760--1772.

\bibitem{42}
D.~Jaschke, M.~L. Wall, L.~D. Carr, Open source matrix product states: Opening
  ways to simulate entangled many-body quantum systems in one dimension,
  Computer Physics Communications 225 (2018) 59--91.

\bibitem{47}
W.-H. Steeb, Y.~Hardy, et~al., Quantum computing and symbolicc++ simulations,
  International Journal of Modern Physics C 11~(2) (2000) 323--334.

\bibitem{5}
C.~C. Moran, Quintuple: A tool for introducing quantum computing into the
  classroom, Frontiers in Physics 6 (2018) 69.

\bibitem{6}
D.~Garc{\'\i}a-Mart{\'\i}n, G.~Sierra, {Five Experimental Tests on the 5-Qubit
  IBM Quantum Computer}, arXiv:1712.05642.

\bibitem{7}
N.~N. Hegade, B.~K. Behera, P.~K. Panigrahi, {Experimental Demonstration of
  Quantum Tunneling in IBM Quantum Computer}, arXiv:1712.07326.

\bibitem{9}
A.~Cervera-Lierta, Exact ising model simulation on a quantum computer,
  arXiv:1807.07112.

\bibitem{10}
P.~J. Coles, S.~Eidenbenz, S.~Pakin, A.~Adedoyin, J.~Ambrosiano, P.~Anisimov,
  W.~Casper, G.~Chennupati, C.~Coffrin, H.~Djidjev, et~al., Quantum algorithm
  implementations for beginners, arXiv:1804.03719.

\bibitem{11}
R.~Harper, S.~Flammia, {Fault tolerance in the IBM Q Experience},
  arXiv:1806.02359.

\bibitem{14}
Z.~Zhao, A.~Pozas-Kerstjens, P.~Rebentrost, P.~Wittek, Bayesian deep learning
  on a quantum computer, arXiv:1806.11463.

\bibitem{28}
K.~Michielsen, M.~Nocon, D.~Willsch, F.~Jin, T.~Lippert, H.~De~Raedt,
  Benchmarking gate-based quantum computers, Computer Physics Communications
  220 (2017) 44--55.

\bibitem{31}
Z.-Y. Han, J.~Wang, H.~Fan, L.~Wang, P.~Zhang, Unsupervised generative modeling
  using matrix product states, Physical Review X 8~(3) (2018) 031012.

\bibitem{17}
A.~S. Bhatia, A.~Kumar, Neurocomputing approach to matrix product state using
  quantum dynamics, Quantum Information Processing 17~(10) (2018) 278.

\bibitem{19}
W.~W. Ho, S.~Choi, H.~Pichler, M.~D. Lukin, Periodic orbits, entanglement and
  quantum many-body scars in constrained models: matrix product state approach,
  arXiv:1807.01815.

\bibitem{22}
I.~L. Markov, Y.~Shi, Simulating quantum computation by contracting tensor
  networks, SIAM Journal on Computing 38~(3) (2008) 963--981.

\bibitem{18}
L.~Bonnes, A.~M. L{\"a}uchli, Superoperators vs. trajectories for matrix
  product state simulations of open quantum system: a case study,
  arXiv:1411.4831.

\bibitem{16}
A.~S. Bhatia, A.~Kumar, Quantifying matrix product state, Quantum Information
  Processing 17~(3) (2018) 41.

\bibitem{23}
E.~Miles~Stoudenmire, D.~J. Schwab, Supervised learning with quantum-inspired
  tensor networks, arXiv:1605.05775.

\bibitem{24}
Y.~Huang, J.~E. Moore, Neural network representation of tensor network and
  chiral states, arXiv:1701.06246.

\bibitem{29}
M.~Dolfi, B.~Bauer, S.~Keller, A.~Kosenkov, T.~Ewart, A.~Kantian, T.~Giamarchi,
  M.~Troyer, {Matrix product state applications for the ALPS project}, Computer
  Physics Communications 185~(12) (2014) 3430--3440.

\bibitem{30}
M.~K. Saggi, S.~Jain, A survey towards an integration of big data analytics to
  big insights for value-creation, Information Processing \& Management 54~(5)
  (2018) 758--790.

\bibitem{32}
P.~A. Zizzi, Theoretical setting of inner reversible quantum measurements,
  Modern Physics Letters A 21 (2006) 2717--2727.

\bibitem{26}
D.~M. Greenberger, {GHZ (Greenberger—Horne—Zeilinger) Theorem and GHZ
  States}, in: Compendium of quantum physics, Springer, 2009, pp. 258--263.

\bibitem{27}
G.~Uchida, {Geometry of GHZ type quantum states}, Ph.D. thesis, Uniwien (2013).

\bibitem{48}
G.-X. Pan, Y.-M. Liu, W.~Zhang, Z.-J. Zhang, Scheme for remotely preparing
  three-particle ghz-class states, International Journal of Modern Physics C
  20~(04) (2009) 557--564.

\end{thebibliography}
\end{document}